\documentclass[preprint,3p, number,twocolumn,sort&compress]{elsarticle}
\usepackage{xcolor}
\usepackage{graphicx}
\usepackage{xurl}
\usepackage{siunitx}
\DeclareSIUnit{\n}{n}
\DeclareSIUnit{\neq}{n_{eq}}
\DeclareSIUnit{\particles}{particles}
\DeclareSIUnit{\cps}{cps}      
\DeclareSIUnit{\Vov}{V_{OV}}
\DeclareSIUnit{\PE}{PE}     
\DeclareSIUnit{\barn}{b}
\DeclareSIUnit{\cm}{\centi\metre}
\DeclareSIUnit{\mm}{\milli\metre}
\DeclareSIUnit{\um}{\micro\metre}
\DeclareSIUnit{\keV}{\kilo\electronvolt}
\DeclareSIUnit{\MeV}{\mega\electronvolt}
\usepackage[version=4]{mhchem}
\usepackage[
    colorlinks=true,
    linkcolor=blue,
    citecolor=blue,
    urlcolor=blue
]{hyperref}
\usepackage{stfloats}

\usepackage{amsmath,amssymb}
\usepackage{booktabs}
\usepackage{multirow}
\usepackage{makecell}
\usepackage{bm}

\begin{document}

\begin{frontmatter}
\title{Radiation Hardness of Commercially Available NUV-MT Silicon Photomultipliers}
\author[a]{T.~Avgitas}
\author[b]{M.~Axiotis}
\author[a]{Z.~Balmforth}
\author[a]{I.~Manthos}
\author[a,c]{K.~Nikolopoulos}
\author[b]{E.~Taimpiri}
\author[a]{C.~Toukmenidis\corref{cor1}}
\ead{christos.toukmenidis@uni-hamburg.de}
\author[b]{A.~Ziagkova}
\cortext[cor1]{Corresponding author}
\affiliation[a]{organization={Institute for Experimental Physics, University of Hamburg},
postcode={22761},
city={Hamburg},
country={Germany}}
\affiliation[b]{organization={NCSR “Demokritos", Institute of Nuclear and Particle Physics},
postcode={153 41},
city={Athens},
country={Greece}}
\affiliation[c]{organization={School of Physics and Astronomy, University of Birmingham},
postcode={B15 2TT},
city={Birmingham},
country={United Kingdom}}

\begin{abstract}
Silicon Photomultipliers (SiPMs) based on the near-ultraviolet, metal-filled trench (NUV-MT) technology offer improved photon detection efficiency and reduced correlated noise relative to earlier designs, making them attractive for a broad range of particle- and astroparticle-physics applications. As such devices may be deployed in high-radiation environments, quantifying their performance after irradiation is essential. Commercially available Broadcom AFBR-S4N series NUV-MT SiPMs were irradiated with \qty{1}{\MeV} neutrons to fluences between \qty{2e9}{} and \qty{1.1e10}{\neq\per\square\cm} at the TANDEM accelerator facility of NCSR ``Demokritos'' in Athens, and characterised before and after irradiation and over successive thermal-annealing stages. Irradiation increased the dark noise by up to three orders of magnitude and degraded the single-photon resolution by up to a factor of ten, with resolution lost entirely at the highest fluence, while no shift in breakdown voltage was observed. Thermal annealing partially recovered the performance, reducing the dark noise by up to a factor of two and restoring the single-photon resolution.
 \end{abstract}

\end{frontmatter}

\section{Introduction}
\label{sec:introduction}
Silicon photomultipliers (SiPMs) have become the photodetector of choice across a growing range of applications including time-of-flight positron emission tomography (TOF-PET)~\cite{Lecoq:2021pet}, high-energy physics (HEP) calorimetry~\cite{CALICE:2022uwn}, and detectors deployed in space-borne missions~\cite{Produit:2023dei,GRID:2022gib} due to their high photon detection efficiency and timing resolution, compactness, low operating voltage, and insensitivity to magnetic fields~\cite{Gundacker:2020cnv, Simon:2018xzl, BISOGNI2019118}. More recently, the near-ultraviolet, metal-filled trench (NUV-MT) technology has further improved the photon detection efficiency in the NUV range and reduced the correlated noise relative to earlier designs, while maintaining high gain and excellent single-photon resolution~\cite{Merzi:2023, Broadcom_AFBR}.
 
In many of these environments, the sensors are exposed to significant particle radiation throughout their operational lifetime, and the resulting damage can substantially degrade their performance. The severity of the radiation field, however, varies enormously between applications. At the extreme, photodetectors foreseen for the high-luminosity upgrades of the LHC experiments must withstand fluences up to \qty{e14}{\particles\per\square\cm}~\cite{Garutti:2019}, well beyond the regime addressed in the present work. Photosensors deployed in scientific space missions and Earth-observation satellites in low-Earth orbit accumulate more moderate fluences, typically in the range $10^{7}$ to $10^{11}$ $\text{n}_{\text{eq}}/\text{cm}^2$ over their operational lifetime~\cite{Merzi:2024}, and it is this regime that is relevant here. Lower still are the fluences encountered in shielded, low-background rare-event searches, for which even the small radiation- and activation-induced increases in dark count 	rate and correlated noise can be of consequence given the stringent noise requirements of such experiments. Understanding and ultimately mitigating radiation damage in SiPMs is therefore essential for their reliable deployment across these environments.
 
Radiation damage in SiPMs has been studied across a range of device technologies and radiation types~\cite{Garutti:2019}. Neutron-irradiation studies have characterised the increase in dark count rate and leakage current associated with bulk displacement damage, extending to very high fluences, up to \qty{5e14}{\neq\per\square\cm}~\cite{CentisVignali:2018}. For NUV-sensitive devices in particular, dedicated campaigns have targeted space applications, quantifying the degradation from both ionising and non-ionising radiation and, in several cases, the stability of the breakdown voltage with fluence~\cite{Merzi:2024, Altamura:2023}. A further consideration for long-term operation is the extent to which this damage can be reversed through thermal annealing, the relation between microscopic defect evolution and the recovery of macroscopic device properties having been established for silicon detectors more generally~\cite{Moll:2002}. For the commercially available NUV-MT devices studied here, however, the radiation response, and in particular its recovery through thermal annealing, has only recently begun to be explored, for example in the context of the CBM RICH detector~\cite{Pena-Rodriguez:2026}, and remains limited. By contrast, the low-temperature and noise performance of this device series prior to irradiation has been characterised in detail~\cite{Avgitas:2026ejp,Niu:2025huc,Liu:2025csi}, providing a well-established non-irradiated  baseline.

In this work, the radiation hardness of commercially available Broadcom AFBR-S4N series NUV-MT SiPMs is investigated. Three AFBR-S4N66P014M devices were irradiated with \qty{1}{\MeV} neutrons to fluences between \qty{2.0e9}{} and \qty{1.1e10}{\neq\per\square\cm} at the TANDEM accelerator facility of NCSR ``Demokritos'' in Athens, Greece, and their key performance parameters, i.e. leakage current, breakdown voltage, single-photon resolution, dark noise, and gain, were characterised before and after irradiation, and over the course of successive thermal-annealing stages. Throughout, the results are compared against the non-irradiated baseline established in a companion characterisation of the same device series~\cite{Avgitas:2026ejp}.
 
This article is organised as follows. Section~\ref{sec:silicon_photomultipliers} describes the operating principles of SiPMs and the mechanisms of radiation-induced damage in silicon. Section~\ref{sec:experimental_setup} details the experimental setup, including the irradiation campaign, the neutron-fluence measurement, the annealing procedure, and the data acquisition and analysis. Section~\ref{sec:results} presents the characterisation results before and after irradiation and throughout annealing, and conclusions are drawn in Section~\ref{sec:conclusions}. 

\section{Silicon Photomultipliers}
\label{sec:silicon_photomultipliers}

This section describes the SiPM as a device: Section~\ref{subsec:sipm_structure} outlines the structure and operating principle of SiPMs and introduces the specific AFBR-S4N model characterised in this work, while Section~\ref{subsec:silicon_damage} summarises the mechanisms by which irradiation damages silicon and degrades SiPM performance.

\subsection{SiPM structure and the AFBR-S4N series}
\label{subsec:sipm_structure}

SiPMs are dense arrays of single photon avalanche photodiodes (SPADs) connected in parallel and operated in Geiger mode, above the breakdown voltage V$_{bd}$. When a single photon is absorbed, a primary electron--hole pair is generated which initiates a self-sustaining avalanche through impact ionisation in the SPAD. The avalanche is subsequently quenched by a resistor connected in series to the SPAD, allowing for the device to recover and maintain stable operation. The total SiPM output is the sum of the responses from all triggered SPADs. 

Each SPAD consists of a thin, heavily doped implant forming a p--n junction with a lightly doped, precisely grown epitaxial layer a few micrometers thick. This epitaxial layer contains the depletion region and, within it, the high-field multiplication region where the Geiger avalanches occur. Beneath the epitaxial layer lies the substrate, a thick (hundreds of micrometers), heavily doped bulk wafer that provides mechanical support and electrical contact to the anode but is not depleted under normal operating bias~\cite{Gundacker:2020cnv}.

A detailed overview of SiPM operating principles, performance parameters, and characterisation methods can be found in Refs.~\cite{Piemonte:2019kll, Klanner:2018ydn, Acerbi:2019qgp,Avgitas:2026ejp}.

The devices characterised in this work are one realisation of this architecture. They are part of Broadcom's near-ultraviolet, metal-filled trench (NUV-MT) SiPM series~\cite{Broadcom_AFBR}, in which metal-filled optical isolation trenches between neighbouring microcells absorb photons emitted during an avalanche before they can propagate, suppressing the internal optical crosstalk relative to earlier designs~\cite{Merzi:2023}. Specifically, the model used is the AFBR-S4N66P014M, a single-channel SiPM with a \qtyproduct{6 x 6}{\mm} active area comprising \num{22428} microcells at a \qty{40}{\um} SPAD pitch. Key performance parameters, as specified by the manufacturer, are summarised in Table~\ref{tab:sipm_specs}.
 
\begin{table}[h!]
    \centering
    \resizebox{\columnwidth}{!}{%
    \begin{tabular}{lc}
        \toprule
        \textbf{Parameter} & \textbf{Value} \\
        \midrule
        Breakdown voltage, $V_{bd}$ & \qty{32.5}{\volt} \\
        Temperature coefficient of $V_{bd}$ & \qty{30}{\milli\volt\per\celsius} \\
        Dark noise ($R_{DN}$) & \qty{125}{\kilo\cps\per\square\mm} \\
        Gain & \num{7.3e6} \\
        Optical crosstalk probability, $p_{CT}$ & \num{0.23} \\
        Afterpulse probability, $p_{AP}$ & $<\num{0.01}$ \\
        Recharge time coefficient & \qty{55}{\nano\second} \\
        \bottomrule
    \end{tabular}%
    }
\vspace{-0.3cm}
    \caption{Manufacturer specification of AFBR-S4N series key performance parameters at \qty{12}{\Vov} and \qty{25}{\celsius}~\cite{Broadcom_AFBR}.\label{tab:sipm_specs}}
\vspace{-0.3cm}
\end{table}

\subsection{Silicon damage due to irradiation}
\label{subsec:silicon_damage}

Radiation damage in silicon sensors depends on both the energy and the type of the radiation, and can be classified as either surface or bulk damage. Bulk damage, caused by non-ionising energy loss (NIEL), results from energetic hadrons (protons, neutrons, pions) and, to a lesser extent, electrons and photons displacing silicon atoms from their lattice sites. A primary knock-on atom leaves behind a vacancy and, if the kinetic energy is sufficient, can initiate a cluster of defects in addition to isolated point defects~\cite{Donegani:2018}. Low-energy electrons mainly produce point defects while protons and neutrons produce both cluster and point defects. Vacancies and interstitials, which are highly mobile within the lattice, can migrate until they annihilate or react with other point defects or impurities present in the silicon, forming stable defect complexes. Such complexes may form either within the dense, disordered environment of a cluster or at sites removed from the original displacement event in the case of isolated point defects. These defect complexes, whether isolated points or clusters, introduce new energy levels within the silicon bandgap that act as generation-recombination centres, trapping centres, or effective dopants, depending on their position in the gap and their occupation~\cite{Lutz:2007}. 

Surface damage, caused by ionising energy loss (IEL), is primarily produced by photons and charged particles depositing energy in the SiO$_2$ passivation layer. Trapped positive charge in the oxide and interface states act as a source of surface generation current, increasing the leakage current below breakdown. Where this current reaches the multiplication region, it is amplified and contributes to the dark count rate~\cite{Garutti:2019}.

Irradiation can also affect the correlated noise of the device, namely afterpulsing and optical crosstalk~\cite{Avgitas:2026ejp}. Afterpulsing occurs when charge carriers generated during an avalanche are captured by trapping centres and subsequently released, following the Shockley--Read--Hall mechanism, triggering a delayed secondary avalanche within the same microcell. As the defect complexes introduced by displacement damage act as additional trapping centres, the afterpulsing probability is expected to increase with fluence~\cite{Garutti:2016hny}. Optical crosstalk, whereby photons emitted during an avalanche trigger discharges in neighbouring microcells, is instead governed by the avalanche charge and the device geometry rather than by the bulk defect density, and is therefore expected to be less sensitive to the displacement damage considered here. 

These microscopic mechanisms result in macroscopically measurable changes in device performance, most notably in an increased leakage current and dark count rate, a degraded single-photon resolution, and a reduced collected signal charge. The evolution of these observables with fluence, and their subsequent recovery through thermal annealing, is the subject of this work.

\section{Experimental setup}
\label{sec:experimental_setup}
The experimental procedure comprises four stages: the neutron irradiation of the devices, the delivered fluence measurement, the thermal-annealing cycles, and the characterisation measurements together with their offline analysis.

\subsection{Irradiation}
\label{subsec:irradiation_setup}

The irradiation was performed at the TANDEM accelerator
facility of NCSR ``Demokritos'' in Athens~\cite{Harissopulos:2021lgh,Lagoyannis:2023mnv}. 
The facility features a \qty{5.5}{\mega\volt} T11/25 Van de Graaff Tandem accelerator hosted in a room of borated concrete walls. 
The beam travels through two experimental areas where both the targets and the experiments are located. 
For high-energy neutron applications, the facility delivers neutrons in the ranges \SIrange{4}{11.5}{\mega\electronvolt}
and \SIrange{16}{20.5}{\mega\electronvolt} via the \ce{^2H(d,n)^3He} and
\ce{^3H(d,n)^4He} reactions, respectively~\cite{Harissopulos:2021lgh}.
However, for the irradiations reported here neutron energies of $\SI{1}{\mega\electronvolt}$ were chosen.
To this end, the \ce{^7Li(p,n)^7Be} reaction on a lithium fluoride (LiF) target is used. Above
its threshold of $E_p = \SI{1.881}{\mega\electronvolt}$, this reaction produces a mono-energetic
neutron group emitted from tens of \si{\kilo\electronvolt} up to several
\si{\mega\electronvolt}, with the neutron energy set by the proton energy and the
emission angle. For proton energies above approximately $\SI{2.37}{\mega\electronvolt}$
a second, lower-energy neutron group appears, corresponding to the
$\SI{429}{\kilo\electronvolt}$ first excited state of \ce{^7Be}. 
At the proton energies required to produce the ground-state neutron group at $\SI{1}{\mega\electronvolt}$,
the first-excited-state channel is also open, producing a second, lower-energy group at approximately \qty{0.5}{\MeV}. 
This second component is estimated to account for approximately \qty{10}{\percent} of the total neutron yield. 
For the purposes of this work, the neutron field can be considered practically mono-energetic. 

\begin{figure}[h!]
    \centering
        \vspace{-0.2cm}
    \includegraphics[width=0.9\linewidth]{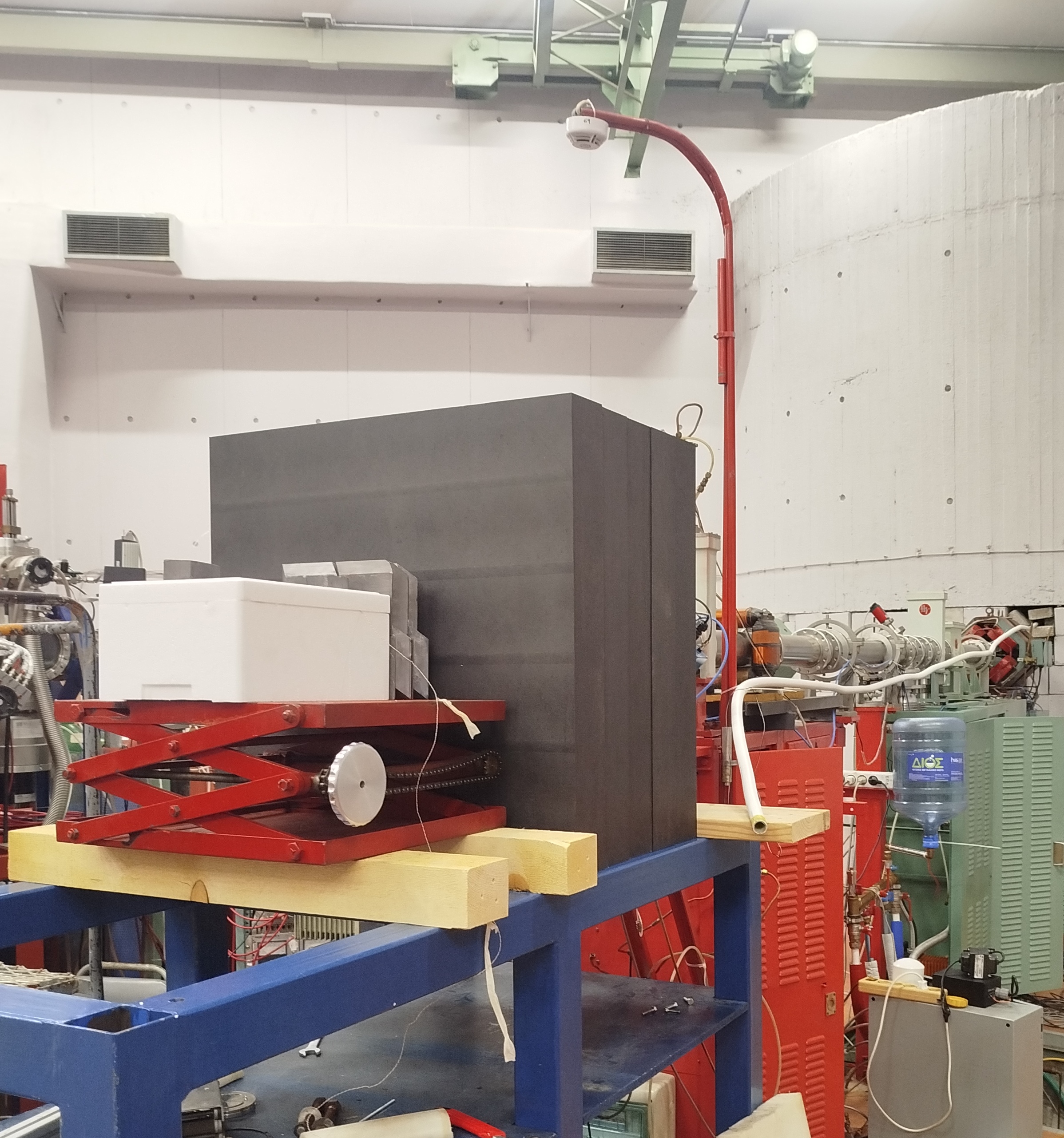}
        \vspace{-0.2cm}
    \caption{SiPM irradiation setup at NCSR ``Demokritos''.\label{fig:setup-1}}
        \vspace{-0.3cm}
\end{figure}
 
Three AFBR-S4N66P014M SiPMs were irradiated with \qty{1}{\MeV} neutrons at a mean flux of approximately \qty{4e5}{\n\per\square\cm\per\second}. As shown in Fig.~\ref{fig:setup-1}, the SiPMs were installed in a thermally insulated box containing dry ice, in order to suppress thermal annealing during irradiation. A temperature of approximately \qty{-35}{\celsius}, monitored at regular intervals with a Type~T Class~1 thermocouple, was maintained throughout the irradiation campaign. To collimate the neutron beam and suppress neutron backgrounds in the experimental hall, blocks of borated high-density polyethylene (HDPE) with a boron content of \SI{10}{\percent} by weight
were placed around the LiF target, with the blocks between the target the devices being irradiated forming a cylindrical channel of \qty{8}{\milli\meter} diameter. To suppress photons produced in the target, lead blocks were positioned between the target and the box. The three SiPMs were placed directly in the neutron beam and were removed at progressively, so that each accumulated a different neutron fluence.
 
\subsection{Fluence Measurements}
\label{subsec:fluence_measurements}

Neutron fluence measurements for the irradiated samples were performed using Indium foils, with a foil placed in front of each SiPM. As summarised in Table~\ref{tab:foils}, two activation pathways of $\mathrm{^{115}In}$ were used, the neutron inelastic scattering and the $(n,\gamma)$ reaction. 
The activity of the post irradiated foil is estimated using one and five de-excitation $\gamma$-ray emission lines, respectively. The measured fluences for each foil, with their uncertainties and the corresponding foil mass, are shown in Table~\ref{tab:fluences}.
 \begin{table}[h!]
    \centering
    \resizebox{\linewidth}{!}{%
    \begin{tabular}{l c c S[table-format=4.1] S[table-format=2.3] S[table-format=1.3(3)]}
        \toprule
        \textbf{Reaction} & \textbf{$t_{1/2}$} & \textbf{$\sigma$ [\unit{\milli\barn}]} & {\textbf{$E_{\gamma}$ [\unit{\kilo\electronvolt}]}} & {\textbf{$I_{\gamma}$ [\%]}} & {\textbf{$\varepsilon$ [\%]}} \\
        \midrule
        $\mathrm{^{115}In}(n,n')\mathrm{^{115m}In}$ & \qty{4.486}{\hour} & 66.2 $\pm$ 2.1 & 336.2 & 45.9 & 0.901 +- 0.130 \\
        \midrule
        \multirow{5}{*}{$\mathrm{^{115}In}(n,\gamma)\mathrm{^{116m}In}$} & \multirow{5}{*}{\qty{54.29}{\minute}} & \multirow{5}{*}{172 $\pm$ 9} & 416.9  & 27.2  & 0.773 +- 0.071 \\
         & & & 818.7  & 12.13 & 0.482 +- 0.043 \\
         & & & 1097.3 & 58.5  & 0.398 +- 0.024 \\
         & & & 1293.6 & 84.8  & 0.360 +- 0.030 \\
         & & & 1507.6 & 9.92  & 0.329 +- 0.046 \\
        \bottomrule
    \end{tabular}%
    }
    \caption{Activation pathways used for the in-situ neutron-fluence measurement. The half-life $t_{1/2}$ and cross-section $\sigma$ are given for each, together with the analysed $\gamma$-ray lines ($E_{\gamma}$), their emission intensity $I_{\gamma}$, and the corresponding photopeak detection efficiency $\varepsilon$. The uncertainties are omitted where negligible. The cross-sections and decay data are taken from the IRDFF-II library~\cite{Trkov:2020irdff}.}
    \label{tab:foils}
\end{table}
 
The two reactions sample the neutron field in complementary ways. The $\mathrm{^{115}In}(n,n')\mathrm{^{115m}In}$ inelastic reaction is a threshold process, with an effective threshold around \qty{0.5}{\MeV} and a cross-section that rises through the \si{\MeV} region, so that it responds practically only  to the \qty{1}{\MeV} group and is practically insensitive to the \qty{0.5}{\MeV} component discussed in Section~\ref{subsec:irradiation_setup}. The $\mathrm{^{115}In}(n,\gamma)\mathrm{^{116m}In}$ reaction, by contrast, has no threshold and a cross-section that is approximately flat across this energy range, so that it responds to both neutron groups. In principle, the two activation measurements together  carry information on the relative contribution of the two groups. In the present setup, however, the associated uncertainties are such that resolving the second group is not possible. 
 
\begin{table}[h!]
    \centering
    \resizebox{\linewidth}{!}{%
    \begin{tabular}{cccc}
        \toprule
        \textbf{SiPM} & \textbf{Mass}& \textbf{Fluence}& \textbf{Uncertainty}\\
          & [\unit{\milli\gram}] & [\unit{\neq\per\square\cm}] &  [$\text{method}\oplus\text{cross-section}$ (\%)] \\
        \midrule
        1 & \num{744.5} & \num{2.0e9} &   11 $\oplus$ 5 \\
        2 & \num{758.8} & \num{4.9e9} &   14 $\oplus$ 5 \\
        3 & \num{167.9} & \num{1.1e10} &  22 $\oplus$ 5 \\
        \bottomrule
    \end{tabular}%
    }
    \caption{Mass of Indium foil used for each SiPM and the corresponding fluence estimates with their uncertainty. The uncertainty is presented as the quadratic sum of a term comprising all experimental and systematic uncertainties of the method and a term representing the uncertainty on the Indium activation reaction cross-section. \label{tab:fluences}}
\end{table}
 
\subsection{Annealing Setup}
\label{subsec:annealing_setup}

Following irradiation, and until the characterisation measurements were performed at the University of Hamburg, the SiPMs were kept in a freezer and were maintained at low temperature during transportation to Hamburg, so that effectively no thermal annealing took place prior to the first post-irradiation characterisation. 
 
The SiPMs were thermally annealed in a temperature-controlled oven. The three devices were placed together inside and their temperature was continuously monitored throughout every cycle. A total of five annealing cycles were performed, following an iterative procedure in which the devices were annealed for a given interval, were characterised, and then annealed further. The first two cycles were performed at temperatures near \qty{60}{\celsius}, while the latter three at temperatures near \qty{75}{\celsius}. The cumulative annealing reached at each stage, as estimated using the actual annealing time and temperatures recorded, is summarised in Table~\ref{tab:annealing_cycles}, expressed in standard units of \qty{80}{\minute} at \qty{60}{\celsius} in accordance with the RD50 collaboration~\cite{Moll:2002,Nikolopoulos:2019wlb}. To calculate the annealing durations, an activation energy of 1.09 $\pm$ 0.14 eV was used for Silicon~\cite{Moll:1999}. The values are cumulative: each characterisation was performed after the total annealing time indicated, so that the annealing-time axis in the following sections corresponds to these cumulative values.

\begin{table}[h!]
    \centering
    \resizebox{0.7\linewidth}{!}{%
    \begin{tabular}{cc}
        \toprule
        \textbf{Annealing}  & \textbf{Cumulative Duration} \\
         \textbf{Cycle} & \textbf{[$\times$~\qty{80}{\minute} at \qty{60}{\celsius}]} \\
        \midrule
        A & 0.89 $\pm$ 0.02 \\
        B & 1.9 $\pm$ 0.1 \\
        C & 4.9 $\pm$ 0.5 \\
        D & 9.5 $\pm$ 1.0 \\
        E & 23.5 $\pm$ 3.6 \\
        \bottomrule
    \end{tabular}%
    }
    \caption{Annealing cycles and their cumulative duration in standard units of \qty{80}{\minute} at \qty{60}{\celsius}. The amount of annealing is estimated using the time and temperatures recorded.
    \label{tab:annealing_cycles}}
\end{table}

\subsection{Data Acquisition and Analysis}
\label{subsec:data_acquisition}

To assess the effect of irradiation and annealing on the SiPM performance, characterisation measurements were conducted in an ISO-7 clean room at the University of Hamburg, using the same setup and analysis as for the previous characterisation of non-irradiated devices of this series~\cite{Avgitas:2026ejp}.
%; the relevant details are summarised here. 
  
% SiPM bias is supplied by a KEITHLEY 2231A-30-3 power supply, with a KEITHLEY 2450 Source Meter Unit used for IV curve acquisitions. The SiPM output signal is amplified using a Mini-Circuits ZFL-1000LN+ low-noise amplifier before being digitised by a Tektronix MSO44B oscilloscope operating at \qty{1}{\giga\hertz} bandwidth and \qty{3.125}{GS/s} sampling rate. An LED with a peak wavelength of \qty{470}{\nano\metre} is mounted inside the dark box and used to illuminate the SiPMs for the IV curve acquisitions. A stable, temperature-controlled environment is achieved using a NORHOF LN$_2$ Microdosing System, which supplies liquid nitrogen to the internal light-tight box and maintains the desired temperature within $\pm\qty{1}{\celsius}$, continuously monitored using a Type~T Class~1 thermocouple and a data logger. All SiPM data were acquired at a temperature of \qty{-30}{\celsius}.
 
The irradiated SiPMs, which are mounted on a custom PCB to provide power and signal readout, are installed inside a light-tight box enclosed within a thermally insulated box and wrapped with a light-tight blanket to achieve the dark conditions necessary for accurate noise rate measurements.
Two acquisition modes are used at each fluence and annealing stage. In the first, a source meter is used, and the breakdown voltage $V_{bd}$ is determined from an IV curve acquired while an LED placed inside the light-tight box is swtiched on: the SiPM bias voltage is incremented and the current drawn is recorded. The $V_{bd}$ is estimated as the voltage at which the derivative of the logarithmic current with respect to voltage, $\mathrm{d\ln I/dV}$, is maximised~\cite{Avgitas:2026ejp}. The operating bias is then set according to $V_{bias} = V_{bd} + V_{OV}$, where $V_{OV}$ is the operating overvoltage.
Subsequently, IV curve data are obtained with the LED switched off, to assess the increase of leakage current following irradiation. 
In the second mode, noise data are acquired with an oscilloscope and the LED switched off: a trigger threshold is set below the single-photoelectron amplitude and individual waveforms are recorded, with a total acquisition window of \qty{200}{\milli\second} per measurement. The data were processed offline to extract the SiPM characterisation parameters. 

\section{Results}
\label{sec:results}
In this section the impact of irradiation and the effect of successive annealing stages on detector performance is examined.

\subsection{Breakdown Voltage}
\label{subsec:vbd}

\begin{figure}[h!]
    \centering
    \vspace{-0.2cm}
        \includegraphics[width=0.95\linewidth]{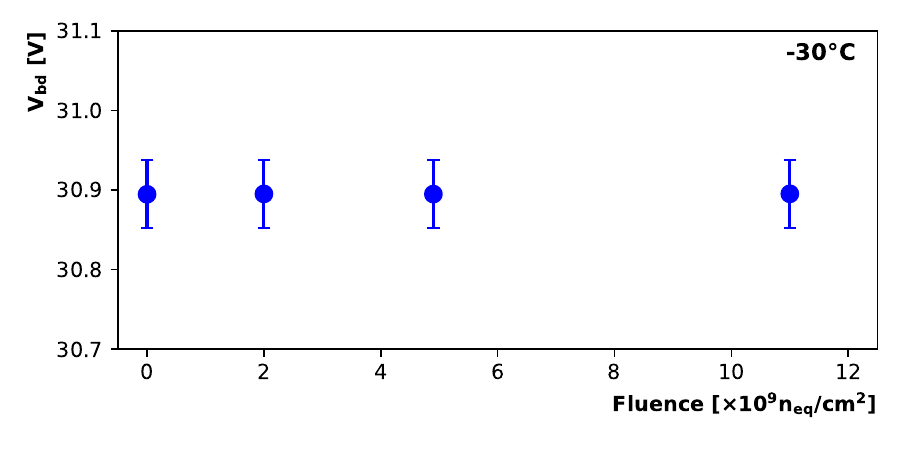}
    \vspace{-0.5cm}
    \caption{Breakdown voltage versus neutron fluence. All measurements were taken at \qty{-30}{\celsius} with the LED on.
    \label{fig:vbd_vs_dose_ledon}}
        \vspace{-0.5cm}
\end{figure}

Breakdown voltage in NUV SiPMs has been shown to remain stable for fluences up to \qty{5.0e11}{\neq\per\square\cm}~\cite{Merzi:2024,Altamura:2023}, indicating that bulk radiation damage from neutron irradiation does not significantly alter the electric field profile or depletion characteristics of the avalanche region. This behaviour is consistent with the results presented here in Fig.~\ref{fig:vbd_vs_dose_ledon}, where no shift in breakdown voltage was observed across the measured fluence range, up to \qty{1.1e10}{\neq\per\square\cm}. Such stability is expected since breakdown voltage is primarily governed by the doping profile and geometry of the junction, both of which are largely unaffected by the displacement damage induced by neutron irradiation at the fluences considered here. 

\subsection{Leakage Current}
\label{subsec:leakage_current}

\begin{figure}[h!]
    \centering
    \includegraphics[width=0.95\linewidth]{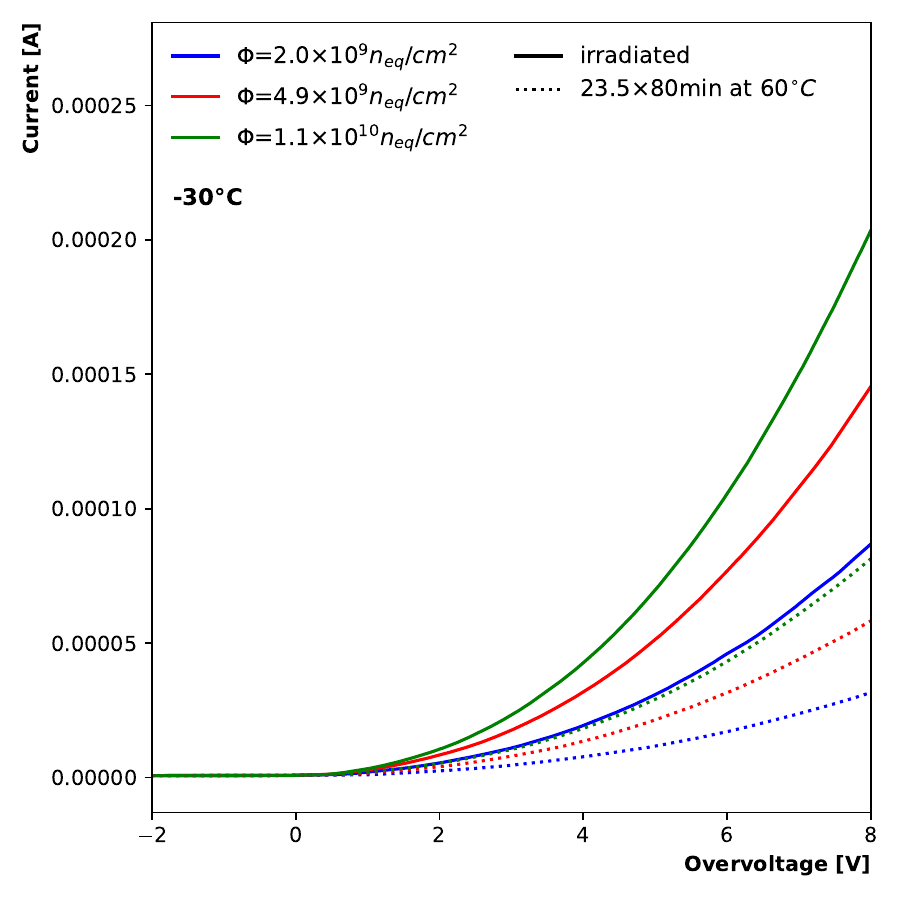}
        \vspace{-0.5cm}
    \caption{IV curves taken with LED switched off for the full fluence range after irradiation (solid lines) and after the 24 annealing units of \qty{80}{\minute} at \qty{60}{\celsius} (dashed lines). All measurements were taken at \qty{-30}{\celsius} with the LED source swtiched off.}
    \label{fig:ivcurves_post_post5anneal_LEDon}
        \vspace{-0.3cm}
\end{figure}

As described in Section~\ref{subsec:silicon_damage}, defects created by irradiation introduce new intermediate states within the silicon bandgap, acting as generation-recombination centres that increase the leakage current through the device. This effect is clearly shown by the solid lines in Fig.~\ref{fig:ivcurves_post_post5anneal_LEDon}, where the leakage current is significantly increased across the full overvoltage range following irradiation. 

Thermal annealing can partially reverse this damage, as elevated temperatures induce the migration and recombination of point defects and the dissociation of unstable complexes, reducing the density of active generation-recombination centres. A decrease in leakage current by a factor of \num{3} is observed at \qty{8}{\Vov} after 24 annealing units of \qty{80}{\minute} at \qty{60}{\celsius}, demonstrating the restorative power of thermal annealing in partially recovering the pre-irradiation performance. However, this recovery is not expected to be complete since defects within cluster cores anneal less readily than isolated point defects, given the high local defect density promotes the re-trapping and reformation of stable complexes~\cite{Moll:2002}. Therefore, a fraction of the radiation-induced defects form stable complexes that are resistant to annealing at these temperatures and timescales. 

\subsection{Single photon resolution}
\label{subsec:spe_resolution}

\begin{figure*}[t]
    \centering
        \vspace{-0.2cm}
    \includegraphics[width=1.\linewidth]{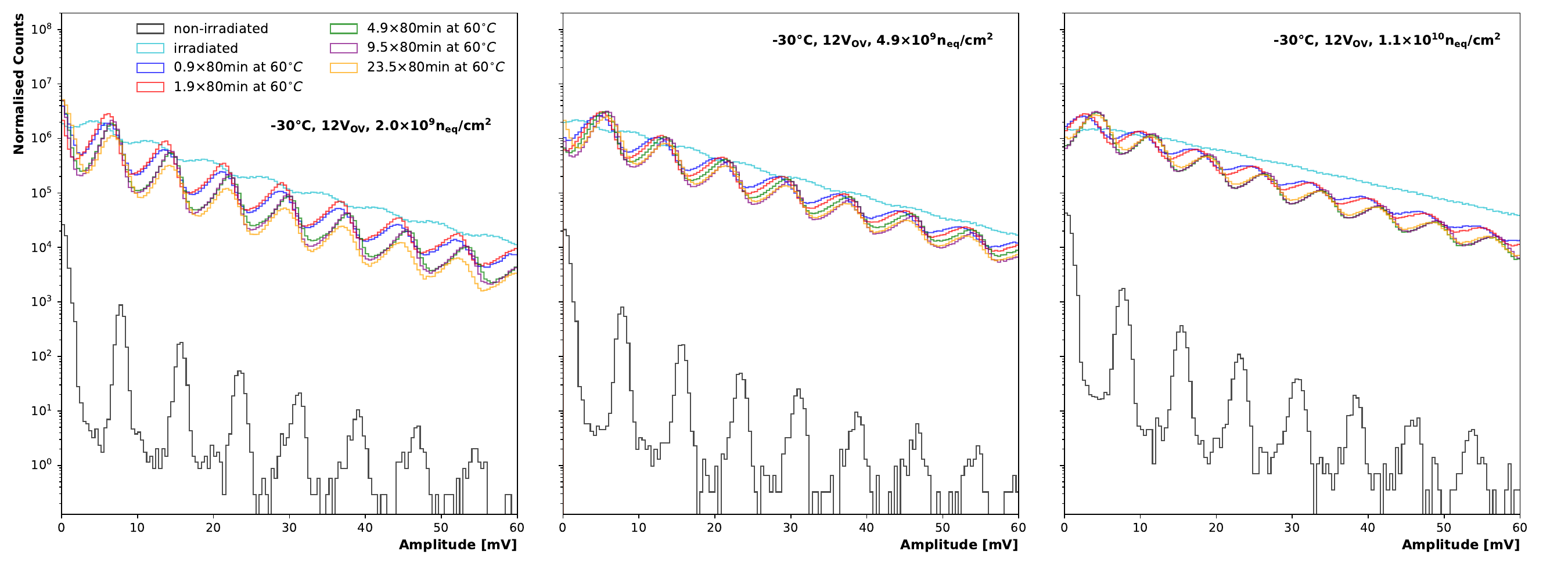}
        \vspace{-0.5cm}
    \caption{Single photoelectron (SPE) amplitude spectra for the three irradiation fluences and the full range of thermal annealing. All measurements were taken at \qty{-30}{\celsius} and \qty{12}{\Vov}.\label{fig:fingerplot_n30deg_12vov_sipm1}}
        \vspace{-0.2cm}
\end{figure*}

The single photon resolution ($R_{spe}$) quantifies the photon-counting capability of a SiPM, characterising how well individual photoelectron (PE) peaks are separated and resolved in the single photoelectron (SPE) amplitude spectrum. 
It is defined as: 
\begin{equation}
    \label{eq:spe_resolution}
    R_{spe} = \frac{\sigma_{1pe}}{\mu_{2pe}-\mu_{1pe}}
\end{equation}
where $\mu$ and $\sigma$ are parameters extracted by fitting the single and two-photon peaks of the SPE spectrum with a Gaussian function.
$R_{spe}$ is degraded by any effect which broadens the single photoelectron peak or reduces the spacing between adjacent photoelectron peaks, such as increased dark count rate, afterpulsing, and optical crosstalk, all of which are known to increase following irradiation as discussed in Section~\ref{subsec:silicon_damage}. Irradiation has a significant effect on $R_{spe}$, while annealing can partially improve it, as shown in Fig.~\ref{fig:fingerplot_n30deg_12vov_sipm1}. 

\begin{figure}[h!]
    \centering
    \vspace{-0.2cm}
    \includegraphics[width=0.95\linewidth]{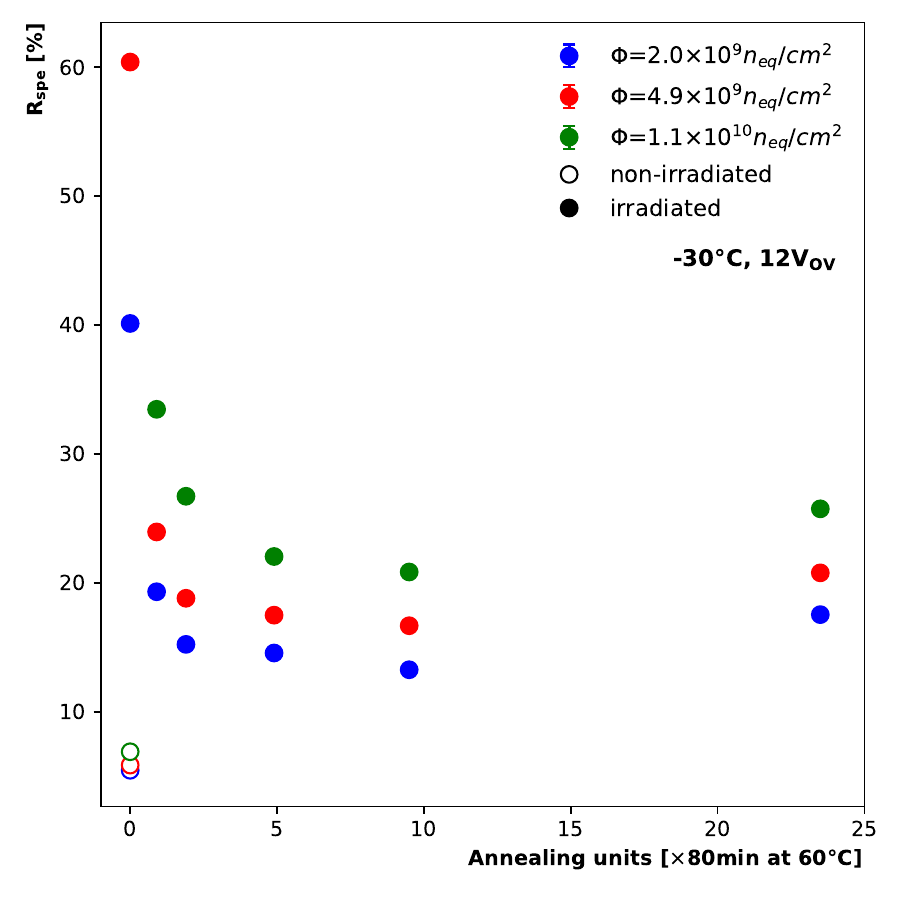}
        \vspace{-0.5cm}
    \caption{Relative single photon resolution defined according to Eq.~\ref{eq:spe_resolution} for the full annealing and fluence ranges. All measurements were taken at \qty{-30}{\celsius} and \qty{12}{\Vov}. Uncertainties in y-axis are included but are not visible.
    \label{fig:spe_resolution}}
        \vspace{-0.3cm}
\end{figure}

The obtained $R_{spe}$  is shown in Fig.~\ref{fig:spe_resolution} across the range of fluences and thermal annealings.
Typical values before irradiation at \qty{-30}{\celsius} and \qty{12}{\Vov} are approximately \qty{6}{\percent}. For irradiated SiPMs, the single photon resolution is degraded significantly, consistent with the increased dark noise and correlated noise contributions. An increase by a factor of \num{7.3}~(\num{10.2}) is observed for \qty{2.0e9}{\neq\per\square\cm}~(\qty{4.9e9}{\neq\per\square\cm}). For \qty{1.1e10}{\neq\per\square\cm}, single photon identification is lost entirely as the elevated noise level broadens and overlaps adjacent photon peaks so they can no longer be individually resolved.

Following annealing, single photon identification is recovered after just 1 annealing unit of 80 minutes at \qty{60}{\celsius}, indicating that a substantial fraction of the damage responsible for the noise increase can be repaired rapidly. After 10 annealing units of 80 minutes at \qty{60}{\celsius}, an improvement by a factor of \num{3.0}~(\num{3.6}) is observed for \qty{2.0e9}{\neq\per\square\cm}~(\qty{4.9e9}{\neq\per\square\cm}), relative to the pre-anneal irradiated values. No further improvement is observed with additional annealing. Instead, a slight degradation is observed. This behaviour is consistent with reverse annealing, whereby the continued evolution of defect complexes at elevated temperature can partially offset the initial recovery~\cite{Moll:2002}. 

Results for the same SiPM technology are reported in \cite{Pena-Rodriguez:2026}, where a factor of \num{5.8} increase in $R_{spe}$ is observed for \qty{1.0e9}{\neq\per\square\cm}, consistent with the trend observed here.

\subsection{Gain}
\label{subsec:gain}

As discussed in Section~\ref{subsec:silicon_damage}, radiation-induced defects act as trapping centres, capturing charge carriers generated during an avalanche and releasing them with a delay. Since the amplitude of a recorded pulse is proportional to the total charge collected within the integration window, this trapping and delayed release reduces the promptly collected signal charge, resulting in an apparent loss of gain despite the underlying avalanche multiplication process being largely unaffected.

This effect can be quantified by the relative gain ($G_{rel}$), defined as follows:

\begin{equation}
    \label{eq:rel_gain}
    G_{rel} = \frac{\mu_{2pe}-\mu_{1pe}}{\mu^{0}_{2pe}-\mu^{0}_{1pe}}
\end{equation}
where $\mu$ denotes the parameters extracted by fitting the single and two-photon peaks of the SPE spectrum in Fig.~\ref{fig:fingerplot_n30deg_12vov_sipm1} with Gaussian functions. The numerator corresponds to the post-irradiation SiPMs, evaluated at each consecutive annealing stage, while the denominator corresponds to the pre-irradiation, reference measurement. $G_{rel} <1$ , therefore, indicates a reduction in collected signal charge relative to the pre-irradiation SiPM, consistent with the increased carrier trapping expected after irradiation.

\begin{figure}[h!]
    \centering   
     \vspace{-0.2cm}
    \includegraphics[width=0.95\linewidth]{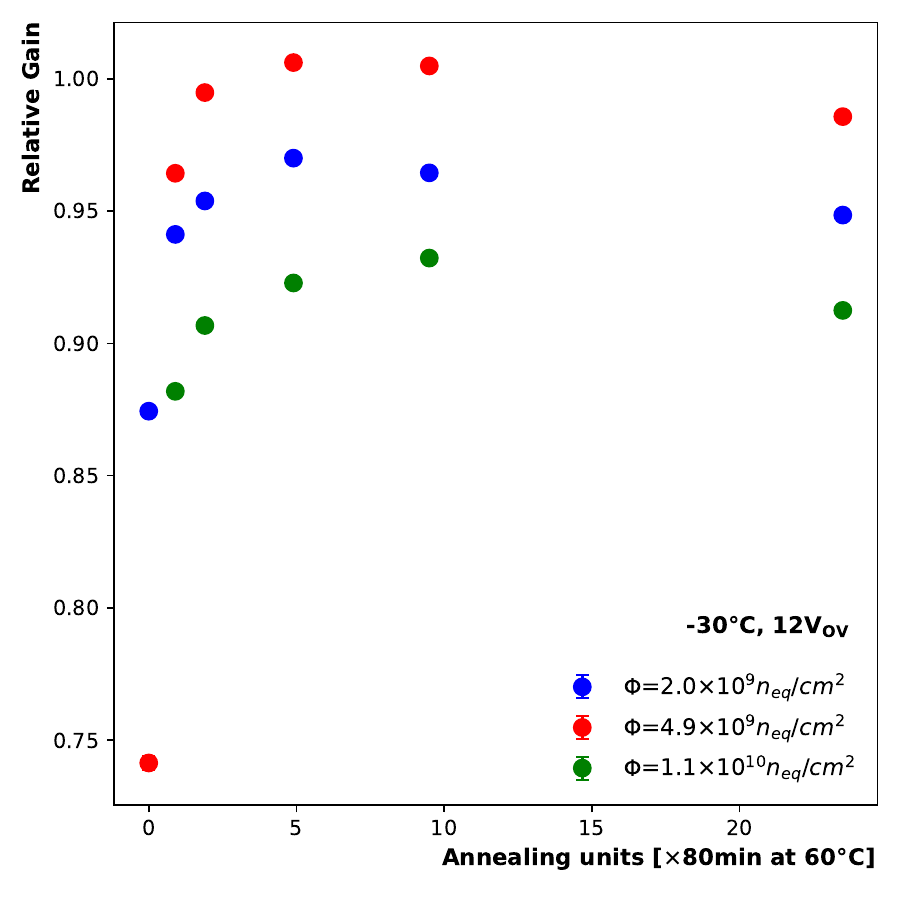}
        \vspace{-0.5cm}
    \caption{Relative gain with respect to pre-irradiation measurements for the full thermal annealing and fluence ranges. All measurements were taken at \qty{-30}{\celsius} and \qty{12}{\Vov}. Uncertainties in are included but are not visible.\label{fig:gain}}
        \vspace{-0.2cm}
\end{figure}

Prior to annealing, all three fluences show a reduction in relative gain.
$G_{rel}$ recovers substantially following thermal annealing for all three fluences, reaching a maximum after approximately 5–10 annealing units of \qty{80}{\minute} at \qty{60}{\celsius}, as shown in Fig.~\ref{fig:gain}. For the lowest two fluences, \qty{2.0e9}{\neq\per\square\cm} and \qty{4.9e9}{\neq\per\square\cm},  $G_{rel}$ recovers to within a few percent of unity, indicating an almost complete recovery of the collected signal charge. For \qty{1.1e10}{\neq\per\square\cm}, the recovery is comparatively smaller, with 
 plateauing at approximately \num{0.93}, reflecting a larger residual population of trapping centres at this fluence. With further annealing beyond 10 units, a slight decrease in $G_{rel}$ is observed across all three fluences, consistent with the reverse-annealing-like behaviour previously discussed~\cite{Moll:2002}.
 
\subsection{Dark Noise}
\label{subsec:dark_noise}

\begin{figure}[h!]
    \centering
        \vspace{-0.2cm}
    \includegraphics[width=0.95\linewidth]{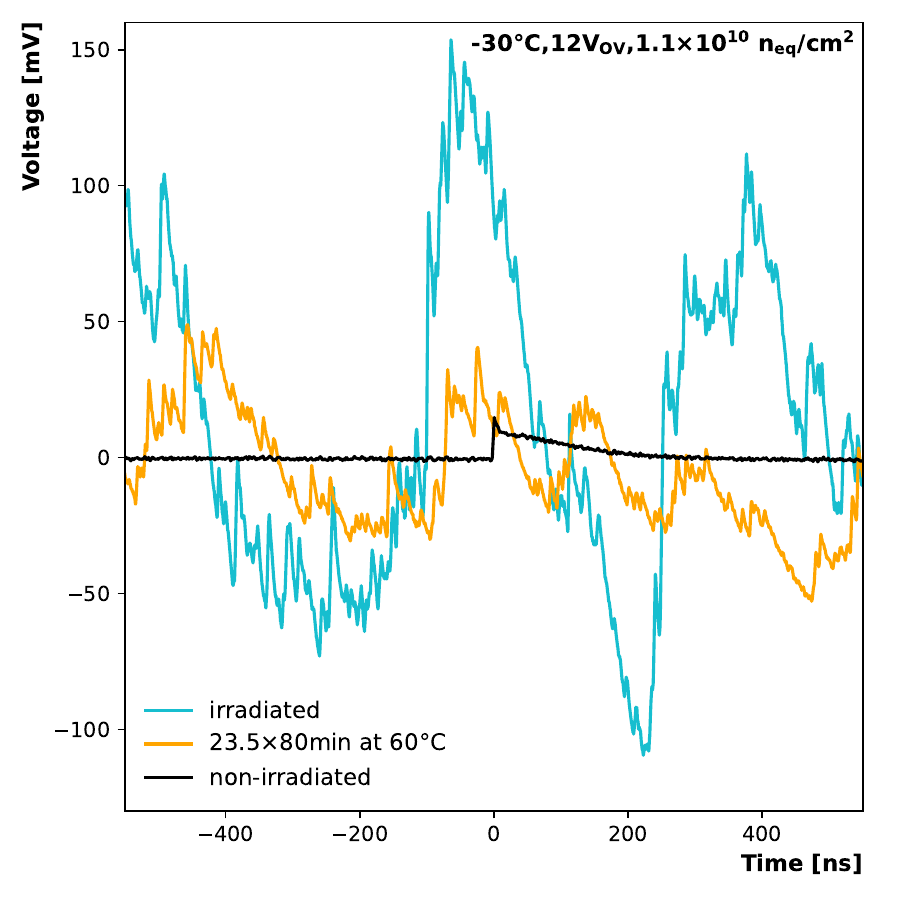}
        \vspace{-0.5cm}
    \caption{Example waveforms before irradiation, after fluence of \qty{1.1e10}{\neq\per\square\cm} and after 24 annealing units of \qty{80}{\minute} at \qty{60}{\celsius}. All waveforms were taken 
    at \qty{-30}{\celsius}, \qty{12}{\Vov}. \label{fig:waveforms_n30deg_12vov_sipm2}}
    \vspace{-0.3cm}
\end{figure}

As discussed in Section~\ref{subsec:silicon_damage}, dark noise is one of the parameters most significantly affected by irradiation. The vacancies, interstitials, and complexes created within the silicon lattice introduce intermediate energy levels within the bandgap that act as generation centres, which can increase the thermally excited carriers in the depletion region even in the absence of incident light. These spurious carriers can trigger avalanche breakdowns in the same way as photon-generated carriers, producing dark count events which are indistinguishable from genuine single-photon signals. This effect is clearly demonstrated in Fig.~\ref{fig:waveforms_n30deg_12vov_sipm2}, where the waveform data are significantly more noisy following irradiation. Annealing can be seen to partially reduce the waveform noise, although full recovery to pre-irradiated status is not achieved within 24 annealing units.

Dark noise is defined as follows:
\begin{equation}
\label{eq:dark_noise}
    R_{DN} = \frac{N_{0.5\,\mathrm{PE}}}{T_{acq}} = R_{0.5\,\mathrm{PE}},
\end{equation}
\noindent where $N_{0.5\,\mathrm{PE}}$ is the total number of pulses with an amplitude exceeding \qty{0.5}{\PE} and $T_{acq}$ is the total acquisition time. 

\begin{figure}[h!]
    \centering
        \vspace{-0.2cm}
    \includegraphics[width=0.95\linewidth]{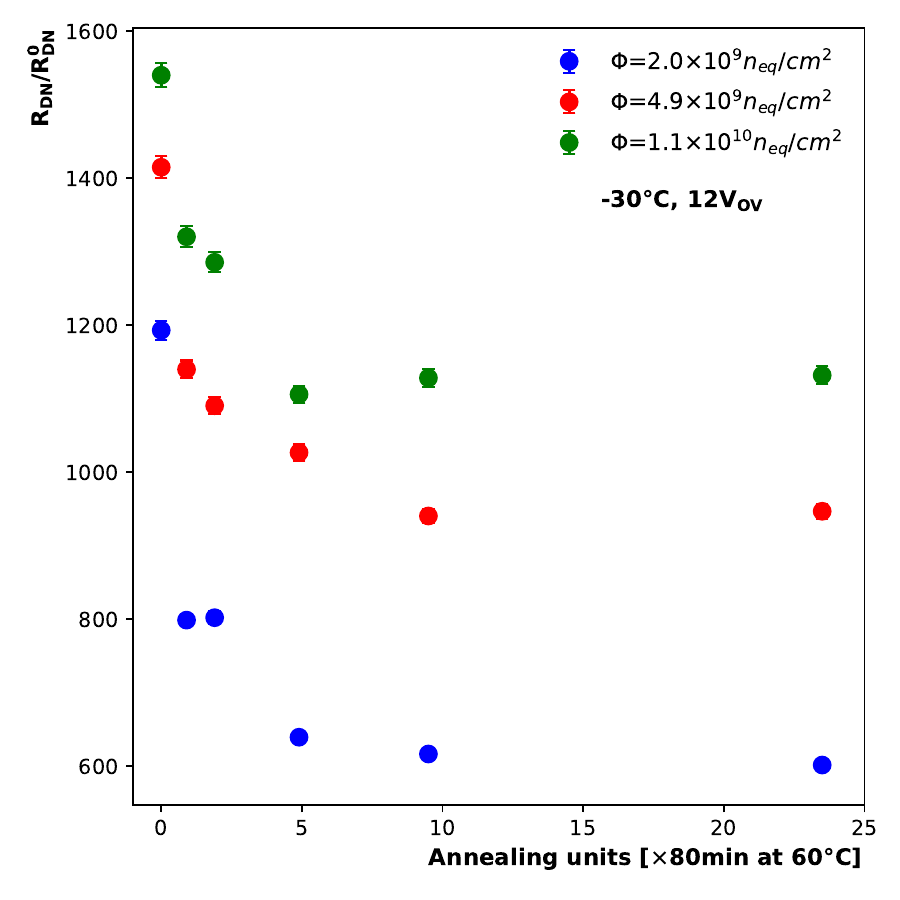}
        \vspace{-0.5cm}
    \caption{Dark noise ratio $R_{DN}/R^0_{DN}$ relative to non-irradiated sample for the full annealing and fluence ranges. All measurements were taken at \qty{-30}{\celsius} and \qty{12}{\Vov}.\label{fig:dcr_vs_annealing_12vov_ratio}}
        \vspace{-0.3cm}
\end{figure}

An increase in dark noise by three orders of magnitude is observed across all three measured fluences, reflecting the substantial density of radiation-induced generation centres introduced even at relatively moderate fluence levels. Although thermal annealing mitigates dark noise, it remains substantially degraded relative to its pre-irradiation level, indicating that a fraction of the generation centres responsible for the dark count rate increase persist even after extended annealing. Following 24 annealing units of \qty{80}{\minute} at \qty{60}{\celsius}, the dark noise is reduced to approximately \num{0.50}, \num{0.67} and \num{0.73} of its post-irradiation value for each respective fluence, as shown in~Fig.~\ref{fig:dcr_vs_annealing_12vov_ratio}. The trend of decreasing recovery fraction with increasing fluence is consistent with a growing contribution from cluster-related defects at higher fluence which, as discussed in Section~\ref{subsec:leakage_current}, are more resistant to thermal annealing than isolated point defects.

\section{Conclusions}
\label{sec:conclusions}
In this work, three AFBR-S4N66P014M NUV-MT SiPMs were irradiated with \qty{1}{\MeV} neutrons to fluences of \qty{2.0e9}{}, \qty{4.9e9}{}, and \qty{1.1e10}{\neq\per\square\cm}, and their key performance parameters were characterised before and after irradiation and over the course of thermal annealing.
 
The breakdown voltage remained stable across the full fluence range, consistent with previous measurements on NUV SiPM technologies~\cite{Merzi:2024,Altamura:2023} and confirming that displacement damage at these fluences does not significantly alter the junction doping profile or geometry. In contrast, the parameters governed by bulk trap and generation-centre density degraded significantly: the leakage current and dark count rate increased substantially at all fluences, the dark count rate by up to three orders of magnitude, while the single-photon resolution degraded by up to a factor of \num{10.2} and was lost entirely at the highest fluence. The relative gain was reduced through increased carrier trapping in the space-charge region.
 
Thermal annealing partially reversed these effects, and was most effective within the first few annealing units. Single-photon identification was recovered after a single annealing unit, and the gain recovered to near-unity for the two lowest fluences. The recovery was, however, incomplete and fluence-dependent, with the recovered fraction decreasing as the fluence increased; while reverse-annealing degradation appeared at later stages. This behaviour is consistent with the growing contribution, at higher fluence, of cluster-related defect complexes, which are more resistant to thermal annealing than isolated point defects~\cite{Moll:2002}.
 
The fluence range investigated here is representative of the cumulative exposure expected for photosensors in scientific space missions and low-Earth-orbit satellites (\qty{e7}{\neq\per\square\cm}--\qty{e11}{\neq\per\square\cm}). Within this range, the stability of the breakdown voltage is favourable for long-term operation, as the operating voltage requires no adjustment over the device lifetime. The substantial and only partially recoverable degradation in dark noise, single-photon resolution, and gain indicates, however, that periodic thermal annealing should be considered as a mitigation strategy where feasible, and that these effects must be incorporated into performance projections for experiments deploying AFBR-S4N66P014M SiPMs in these radiation environments.

\section*{Acknowledgements}
The support of the Deutsche Forschungsgemeinschaft (DFG, German Research Foundation) under Germany’s Excellence Strategy – EXC 2121 “Quantum Universe”-390833306 is acknowledged. 
The authors are grateful to the technical and operations staff of the Tandem Accelerator Laboratory at NCSR Demokritos for their support during the measurements.

\bibliographystyle{elsarticle-num}
\bibliography{references}   

\end{document}